\documentclass[%
 reprint,
 amsmath,amssymb,
 aps,
]{revtex4-2}

\usepackage{graphicx}% Include figure files
\usepackage{dcolumn}% Align table columns on decimal point
\usepackage{bm}% bold math
\begin{document}

\preprint{APS/123-QED}

\title{Photon-assisted entanglement and squeezing generation and decoherence suppression via a quadratic optomechanical coupling}
\author{Zhucheng Zhang}
\author{Xiaoguang Wang}
\email{xgwang1208@zju.edu.cn}
\affiliation{Zhejiang Institute of Modern Physics, Department of Physics, Zhejiang University, HangZhou 310027, China}
\date{\today}

\begin{abstract}
Entanglement and quantum squeezing have wide applications in quantum technologies due to
their non-classical characteristics. Here we study entanglement and quantum squeezing in an
open spin-optomechanical system, in which a Rabi model (a spin coupled to the mechanical oscillator)
is coupled to an ancillary cavity field via a quadratic optomechanical coupling.
We find that their performances can be significantly modulated via the photon of the ancillary
cavity, which comes from photon-dependent spin-oscillator coupling and detuning. Specifically, a
fully switchable spin-oscillator entanglement can be achieved, meanwhile a strong mechanical
squeezing is also realized. Moreover, we study the environment-induced decoherence and
dissipation, and find that they can be mitigated by increasing the number of photons. 
This work provides an effective way to manipulate entanglement and quantum squeezing and to suppress 
decoherence in the cavity quantum electrodynamics with a quadratic optomechanics.
\end{abstract}
\maketitle

\section{\label{sec:level1}Introduction}

Entanglement and quantum squeezing, as fascinating quantum effects, are important
resources in quantum technologies, such as quantum information
\cite{RevModPhys.77.513,RevModPhys.81.865}, quantum computing \cite{divincenzo1995quantum},
quantum metrology \cite{eberle2010quantum,braginsky1995}, and so on. Entanglement
characterizes the correlations between observables that cannot be understood with the local realistic theories \cite{bell1964physics}, which
arouses great attention of many researchers. For example, entanglement has been realized 
in experiments with microscopic systems including atoms \cite{PhysRevLett.79.1,julsgaard2001,su2014}, ions \cite{turchette1998,kielpinski2002} and 
photons \cite{kwiat1995,wang2009linear}. For macroscopic
systems, entanglement between an optical field and a macroscopic vibrating mirror has been
shown to be generated by radiation pressure \cite{vitali2007optomechanical,genes2008robust,yan2017enhanced,deng2016optimizing,wang2015bipartite,barzanjeh2011entangling,kuzyk2013generating,liao2018reservoir}. Meanwhile, entanglement
between different mechanical oscillators has also been studied in various optomechanical
systems \cite{huang2009entangling,liao2014entangling,chen2014enhancement,wang2016macroscopic,li2017enhanced,chakraborty2018entanglement}. Quantum squeezing is instead potentially useful for surpassing the quantum noise limit \cite{braginsky1995}. Many researchers design various systems to realize quantum squeezing. For example,
researchers have obtained strong quadrature squeezing in a transparent crystal with a $\chi^{(2)}$
or a $\chi^{(3)}$ nonlinear polarization \cite{wu1986generation,hilico1992squeezing}. What's more, an optomechanical system, in its steady state, was shown can mimic a medium with $\chi^{(3)}$ nonlinearities \cite{fabre1994quantum,mancini1994quantum}, which
motivates researchers to design lots of optomechanical systems to generate quantum 
squeezing of optical and mechanical modes \cite{sete2010interaction,marino2010classical,brooks2012non,jahne2009cavity,almog2007noise,zhang2018quantum}. To realize entanglement and quantum squeezing
in macroscopic system has always been the focus of research, and the optomechanical
system is undoubtedly a promising research platform.

Cavity quantum electrodynamics (QED) aims to study the quantum behavior of atoms (ions)
confined in a specific space interacting with light fields \cite{walther2006cavity}. In order to observe more abundant physical phenomena, researchers have been working to expand research platforms. For example, a strong coupling between mechanical oscillator and atom has been investigated in the cavity QED combined with a linear optomechanicas \cite{hammerer2009strong}. Recently, a quadratic optomechanics was introduced into
the cavity QED to study the superradiant quantum phase transition \cite{lu2018single,lu2018entanglement}, which was found that the
realized quantum phase transition can be immune to the no-go theorem. Besides, a fully switchable
phonon blockade was also realized in this type of system \cite{zheng2019single}. In the quadratic optomechanics, the phonon
potential can be modulated by the photon through the quadratic optomechanical coupling \cite{bhattacharya2008optomechanical,thompson2008strong,sankey2010strong}, which
makes it possible to use photon to manipulate the properties of system. Thus, an interesting
question is whether one can use photon to manipulate the entanglement or the quantum squeezing
in the cavity QED combined with a quadratic optomechanics.

In order to investigate the entanglement and the quantum squeezing in the cavity QED combined
with a quadratic optomechanics, here we consider a hybrid quantum system, i.e.,
a Rabi model (a spin coupled to the mechanical oscillator) coupled to an ancillary cavity
field via a quadratic optomechanical coupling. In this hybrid system, photon-assisted spin-oscillator entanglement
and mechanical squeezing are realized. Specifically, through controlling the number of photons in
the auxiliary cavity, a fully switchable spin-oscillator entanglement can be achieved, meanwhile a
strong mechanical squeezing is also realized. Moreover, the effects of the environment-induced
decoherence and dissipation on the system can be mitigated by increasing the number of
photons. 

We should point out that the entanglement in the cavity QED combined with the quadratic
optomechanics was studied in Ref.~\cite{lu2018entanglement}, which is associated with 
a single-photon-induced quantum phase transition. But the entanglement realized in their scheme needs an extremely strong quadratic
optomechanical coupling (about a quarter of mechanical frequency), which still exists an experimental challenge. 
However, in our paper, we study entanglement and quantum squeezing in the open quantum model, and the realized 
entanglement and quantum squeezing come from photon-dependent spin-oscillator coupling and detuning. What's more, with the same system,
the parameters used in our scheme have more potential for experimental implementation. It is fundamental important to realize fully switchable
entanglement and strong quantum squeezing in the open macro-system, which
should have wide applications in the field of modern quantum technologies.

The paper is organized as follows:  In Sec.~\uppercase\expandafter{\romannumeral2}, 
we describe the hybrid quantum model and
analyze its experimental feasibility. In Sec.~\uppercase\expandafter{\romannumeral3} and 
in Sec.~\uppercase\expandafter{\romannumeral4}, we derive photon-assisted
spin-oscillator entanglement and mechanical squeezing, respectively. In addition, the effects
of the environment-induced decoherence and dissipation on the system are also considered.
Finally, we summarize our main results in Sec.~\uppercase\expandafter{\romannumeral5}.

\section{\label{sec:level2}Model and Hamiltonian}

\begin{figure}[h!]
\centering\includegraphics[width=4cm]{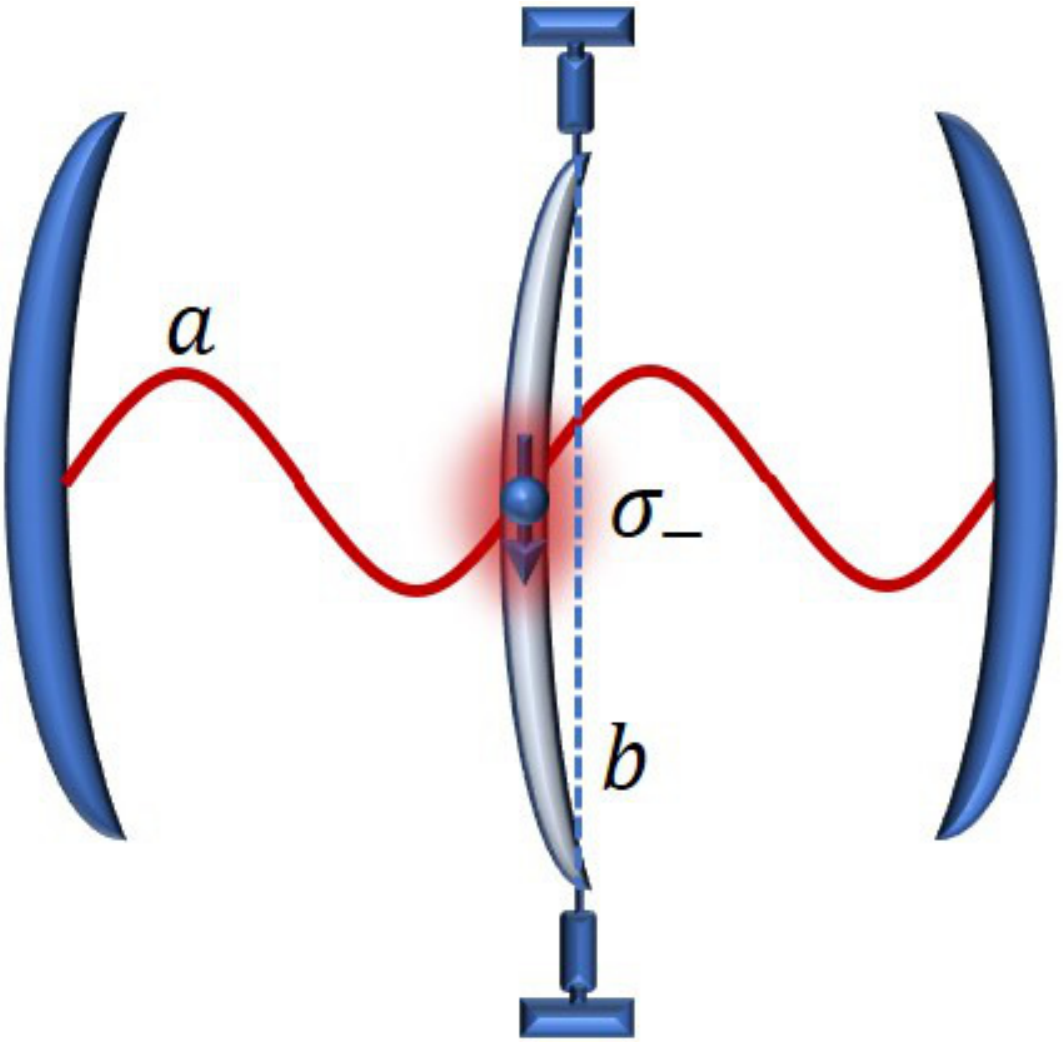}
\caption{Schematic diagram of the system. A two-level system $\sigma_{-}$ (e.g.,
nitrogen-vacancy (NV) center spin) is coupled to a mechanical oscillator $b$. In addition, 
the mechanical oscillator is located in the node (or antinode) of the intracavity field $a$.}
\end{figure}

As shown in Fig.~1, we consider a hybrid optomechanical system, in which a mechanical
oscillator is located in the node (or antinode) of the intracavity field, which can generate
a quadratic optomechanical coupling \cite{bhattacharya2008optomechanical,thompson2008strong,sankey2010strong}.
Besides, a two-level system is coupled to the
mechanical oscillator (e.g., a nitrogen-vacancy (NV) center spin embedded in diamond),
which ultimately constitutes the Rabi model \cite{arcizet2011single,bennett2013phonon,teissier2014strain}. The total Hamiltonian of the system can be
written as ($\hbar=1$),
\begin{equation}
H=H_{an}+H_{rm}-ga^{\dagger}a\left(b^{\dagger}+b\right)^{2},
\end{equation}
with
\begin{align}
H_{an}=&\omega_{a}a^{\dagger}a,\\
H_{rm}=&(\Omega/2)\sigma_{z}+\omega_{b} b^{\dagger}b+\lambda\left(b^{\dagger}+b\right)\sigma_{x},
\end{align}
in which $a (a^{\dagger})$ and $b (b^{\dagger})$ denote the annihilation (creation) 
operators of the cavity (with resonant frequency $\omega_{a}$) and mechanical
(with resonant frequency $\omega_{b}$) modes. $g$ characterizes the quadratic optomechanical
coupling. $H_{an}$ and $H_{rm}$ describe the Hamiltonian of the cavity field and the Rabi model,
respectively. $\sigma_{z}$ and $\sigma_{x}$ denote Pauli operators of the spin (with transition
frequency $\Omega$), and $\lambda$ is the coupling strength between the spin and the oscillator.

From Eq.~(1), one can find that the potential of oscillator is dependent on the intracavity photon
number, so we can manipulate the properties of system via controlling the photon number.
In other words, the cavity field $a$ can be seen as an ancillary field, and in the following, we
assume that the intracavity field $a$ is prepared into the Fock state 
$\left| n\right\rangle  (n=1,~2,~3...)$.  Based on our scheme,  the entanglement and the quantum
squeezing are significantly manipulated by the photon as shown in the following sections.
Through projecting the system Hamiltonian into the cavity field (Fock state) subspace, and then
applying a squeezing transformation with squeezing operator $S(r_{n})=$exp$[r_{n}(b^{2}-b^{\dagger2})]$  (squeezing amplitude $r_{n}=-\frac{1}{4}\ln\left(1-4ng/\omega_{b}\right)$), Eq.~(1) can be
simplified as,
\begin{align}
H_{eff} & =S(r_{n})HS^{\dagger}(r_{n}) \notag \\
&=\frac{\Omega}{2}\sigma_{z}+\omega_{n}b^{\dagger}b+\lambda_{n}\left(b^{\dagger}+b\right)\sigma_{x},
\end{align}
in which the constant term has been neglected, $\omega_{n}=\exp\left(-2r_{n}\right)\omega_{b}$
and $\lambda_{n}=\exp\left(r_{n}\right)\lambda$ are transformed mechanical frequency and 
spin-oscillator coupling strength, respectively. 
\begin{figure}[h!]
\centering\includegraphics[width=7.5cm]{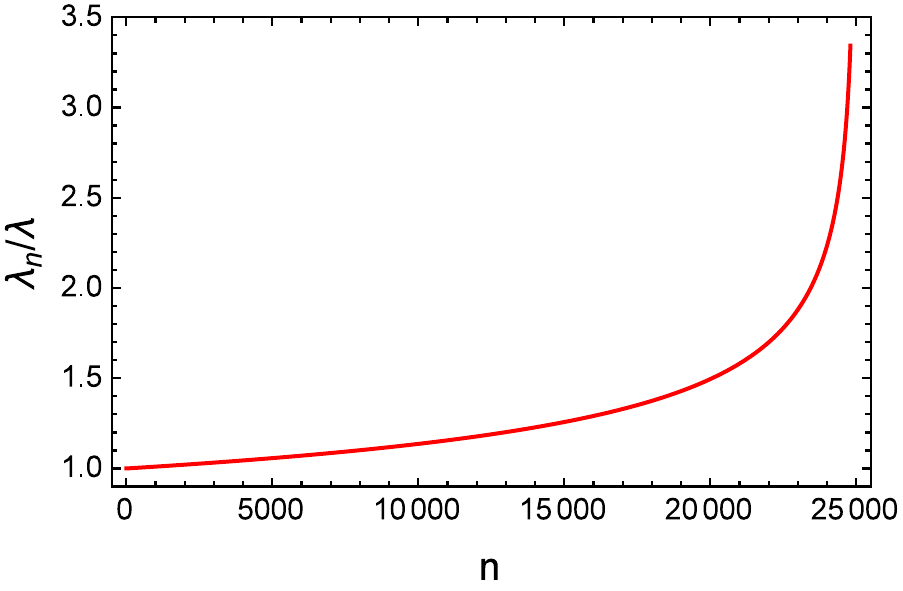}
\caption{Plot of the transformed spin-oscillator coupling strength $\lambda_{n}$ as a function of the
photon number $n$. The parameters are $\lambda=\kappa$
($\kappa$ is the decay rate of the oscillator), $\omega_{b}=2000\kappa$ and
$g=10^{-5}\omega_{b}$.}
\end{figure}

As shown in Fig.~2, the transformed spin-oscillator coupling strength $\lambda_{n}$ is plotted
as a function of the photon number $n$. One can find that the transformed
spin-oscillator coupling can be effectively enhanced with the increase of the photon. Here we
should point out that, in our selected system parameters, the quadratic optomechanical
coupling $g$ is chosen to be $10^{-5}\omega_{b}$, which might be achieved in
the optomechanical system by significantly
enhancing the quadratic optomechanical coupling, and there are many theoretical and experimental
schemes, such as near-field effects \cite{li2012proposal}, fiber-cavity-based optomechanical device
\cite{flowers2012fiber} and so on. Besides, with the superconducting
circuit \cite{kim2015circuit}, our scheme have more potential for experimental implementation than 
Ref.~\cite{lu2018entanglement}.
We should also point out that the enhancement of the spin-oscillator coupling dose not break the
condition for the rotating wave approximation (RWA). Specifically, 
when the photon number $n=20000$, the ratio $\omega_{n}/ \lambda_{n} \approx 598$; 
when the photon number $n=24000$, the ratio $\omega_{n}/ \lambda_{n} \approx 179$.
That is say, the RWA can still be adopted, and the effects caused by the anti-rotating wave term
can be neglected. Thus, under the RWA, Eq.~(4) in the interaction picture can be written as,
\begin{equation}
H_{I}=\lambda_{n}\left(b^{\dagger}\sigma_{-}e^{-i\Delta t}+b\sigma_{+}e^{i\Delta t}\right),
\end{equation}
in which $\Delta=\Omega-\omega_{n}$ is the photon-dependent detuning between the spin 
and the oscillator. One can find that the above Hamiltonian is actually a photon-dependent
Jaynes-Cummings model.

\section{\label{sec:level3}photon-assisted spin-oscillator entanglement}
In this section, we investigate the spin-oscillator entanglement. We consider that the initial state of
the system is $|\psi(0)\rangle=|\uparrow, 0\rangle$, that is, the spin is in the spin-up state
$|\uparrow\rangle$; and the oscillator is prepared into the ground state $|0\rangle$, which can be
realized with the optical back action \cite{poggio2007feedback,bhattacharya2007trapping}. One can find that with the interection Hamiltonian shown in
Eq.~(5), there is only the transition, i.e., $|\uparrow, 0\rangle \leftrightarrow |\downarrow, 1\rangle$.
Thus,  the system state at time $t_1$ can be assumed to be
\begin{equation}
|\psi(t_{1})\rangle 
=c_{\uparrow,0}(t_{1})|\uparrow,0\rangle+c_{\downarrow,1}(t_{1})|\downarrow,1\rangle,
\end{equation}
with the probability amplitudes $c_{\uparrow,0}(t_{1})$ and $c_{\downarrow,1}(t_{1})$.
By substituting Eqs.~(5)-(6) into the Schr{\"o}dinger equation (see appendix for details), we can get the probability amplitudes
as,
\begin{align}
c_{\uparrow,0}(t_{1}) & =\left[\cos\left(\frac{\widetilde{\Omega}_{n}}{2}t_{1}\right)-i\frac{\Delta}{\widetilde{\Omega}_{n}}\sin\left(\frac{\widetilde{\Omega}_{n}}{2}t_{1}\right)\right]\exp(i\Delta t_{1}/2),\\
c_{\downarrow,1}(t_{1}) & =-i\frac{\Omega_{n}}{\widetilde{\Omega}_{n}}\sin\left(\frac{\widetilde{\Omega}_{n}}{2}t_{1}\right)\exp(-i\Delta t_{1}/2),
\end{align}
in which $\Omega_{n}=2\lambda_{n}$ can be seen as a photon-dependent quantum Rabi
frequency, and $\widetilde{\Omega}_{n}=\sqrt{\Omega_{n}^{2}+\Delta^2}$ is rescaled quantum Rabi
frequency due to the detuning $\Delta$.

From the system state at $t_1$, one can find that after the evolution, the oscillator and the spin
evolve from a product state to an entangled state. The entangled state is a two-mode pure state,
which can be measured by the concurrence \cite{wootters1998entanglement}. Based on the definition of concurrence for pure state,
we can get the concurrence at $t_1$ as
\begin{equation}
C=2\left|c_{\uparrow,0}(t_{1})c_{\downarrow,1}(t_{1})\right|.
\end{equation}

\begin{figure}[h!]
\centering\includegraphics[width=7.5cm]{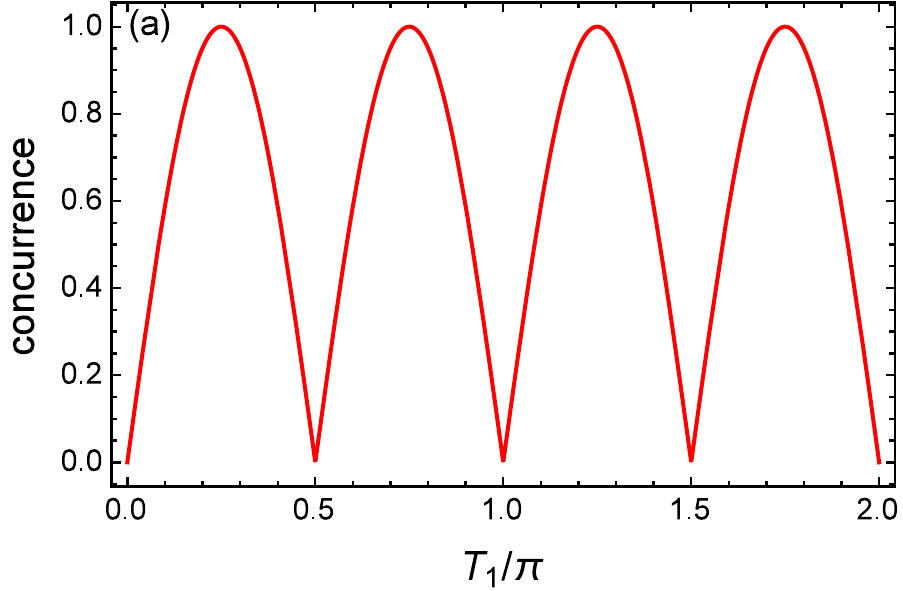}
\centering\includegraphics[width=7.5cm]{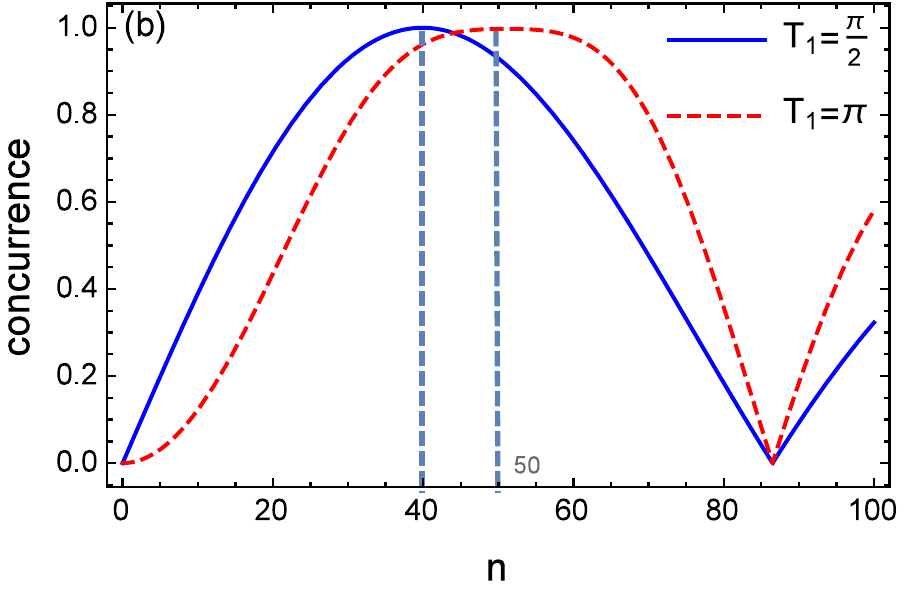}
\caption{The concurrence for the system state is plotted as a function of (a) the scaled
time $T_1$ and (b) the photon number $n$, respectively. 
The parameters are (a) $n=0$, $\Omega=\omega_b$, $T_{1}=\kappa t_1$, (b) $T_{1}=\pi/2, \pi$,
and other parameters are the same as in Fig.~2. }
\end{figure}
As shown in Fig.~3, the concurrence of the system state is plotted as a function of the time
and the photon number, respectively. From Fig.~3(a), one can observe that, when there are no
photons in the ancillary cavity, the concurrence evolves with time periodically, and can reach to the
maximum entanglement, i.e., the concurrence $C=1$. In addition, at time 
$T_{1}=\frac{\pi}{2}j (j=0,~1,~2,...)$, the curve of concurrence is at
the bottom of the valley, i.e., there is no entanglement between the spin and the oscillator.
However, when we inject photons into the auxiliary cavity at these specific moments 
(e.g., $T_{1}=\pi/2,~\pi$), one can find from Fig.~3(b) that a fully switchable spin-oscillator
entanglement can be achieved. In other words, through reasonably controlling the photon number
in the auxiliary cavity, we can in principle achieve a high (without the appearance of the valley)
entanglement degree between the spin and the oscillator in the evolution of time. 

This manipulatable spin-oscillator entanglement comes from photon-dependent spin-oscillator
coupling $\lambda_{n}$ and detuning $\Delta$, as shown in Eq.~(5). Specifically, Due to that the
spin-oscillator coupling and the dutuning can be adjusted by the photon number, the quantum Rabi
frequency $\Omega_{n}$ is also dependent on the photon number, which results that the period of 
entanglement oscillation in evolution time can be manipulated. Hence, we can
control the entanglement degree, as shown in Fig.~3(b).

We should point out that up to now, we have not considered the effects of environment-induced
decoherence and dissipation on the system. In practice, these effects are not negligible factors.
Now we investigate the environment-induced decoherence and dissipation after the entangled
state $|\psi(t_{1})\rangle$ has been prepared. We consider that the oscillator is in the thermal bath
with thermal phonon number $n_{th}$, then the dynamics of system can be described by the 
following master equation,
\begin{align}
\frac{d\rho}{dt}&=-i[H_{I},\rho]+\frac{\text{\ensuremath{\kappa}}}{2}(n_{th}+1)(2b\rho b^{\dagger}-b^{\dagger}b\rho-\rho b^{\dagger}b)  \notag \\
&+\frac{\kappa}{2}n_{th}(2b^{\dagger}\rho b-bb^{\dagger}\rho-\rho bb^{\dagger}) \notag \\
&+\frac{\gamma_{a}}{2}(2\sigma_{-}\rho\sigma_{+}-\sigma_{+}\sigma_{-}\rho-\rho\sigma_{+}\sigma_{-}),
\end{align} 
with the decay rates for oscillator ($\kappa$) and spin ($\gamma_{a}$). 

When the environment-induced decoherence and dissipation are considered, the system state
will no longer be a pure state. The density matrix $\rho(t)$ in the basis 
($|1\rangle=|\uparrow,0\rangle$, $|2\rangle=|\uparrow,1\rangle$,
$|3\rangle=|\downarrow,0\rangle$, $|4\rangle=|\downarrow,1\rangle$) can be written as
\begin{equation}
\left(\begin{array}{cccc}
\rho_{11} & 0 & 0 & \rho_{14}\\
0 & 0 & 0 & 0\\
0 & 0 & \rho_{33} & 0\\
\rho_{41} & 0 & 0 & \rho_{44}
\end{array}\right).
\end{equation}
Substituting Eq.~(11) into the master equation, one can get the time evolution of the matrix elements
as follows,
\begin{align}
\frac{d\rho_{11}}{dt}&=-i\lambda_{n}(e^{i\Delta t}\rho_{41}-e^{-i\Delta t}\rho_{14})-\gamma_{a}\rho_{11},\\
\frac{d\rho_{14}}{dt}&=-i\lambda_{n} e^{i\Delta t} (\rho_{44}-\rho_{11})-\frac{\kappa(n_{th}+1)+\gamma_{a}}{2}\rho_{14},\\
\frac{d\rho_{33}}{dt}&=\kappa(n_{th}+1)\rho_{44}-\kappa n_{th}\rho_{33}+\gamma_{a}\rho_{11},\\
\frac{d\rho_{41}}{dt}&=i\lambda_{n} e^{-i\Delta t} (\rho_{44}-\rho_{11})-\frac{\kappa(n_{th}+1)+\gamma_{a}}{2}\rho_{41},\\
\frac{d\rho_{44}}{dt}&=i\lambda_{n}(e^{i\Delta t}\rho_{41}-e^{-i\Delta t}\rho_{14})-\kappa(n_{th}+1)\rho_{44}+\kappa n_{th}\rho_{33}.
\end{align}
The above differential equations can be solved numerically, then the density matrix $\rho(t)$ can
be gotten.

From Eqs.~(12)-(16), one can see that the density matrix of system will decay in the evolution with
time due to the interaction between system and environment, in which the decay of diagonal
moments corresponds to the loss of the system energy, while the decay of non-diagonal elements
often accompanies the decay of quantum coherence \cite{scully1997quantum}. Decoherence has
always been a very important problem in quantum optics and quantum information. In order to
investigate the role of photon of the auxiliary cavity in the interaction between system and
environment, we numerically simulate the time evolution of
diagonal and non-diagonal elements of the system density matrix $\rho(t)$ under different
photon numbers, respectively, as shown in Fig.~4.
\begin{figure}[h!]
\centering\includegraphics[width=7.5cm]{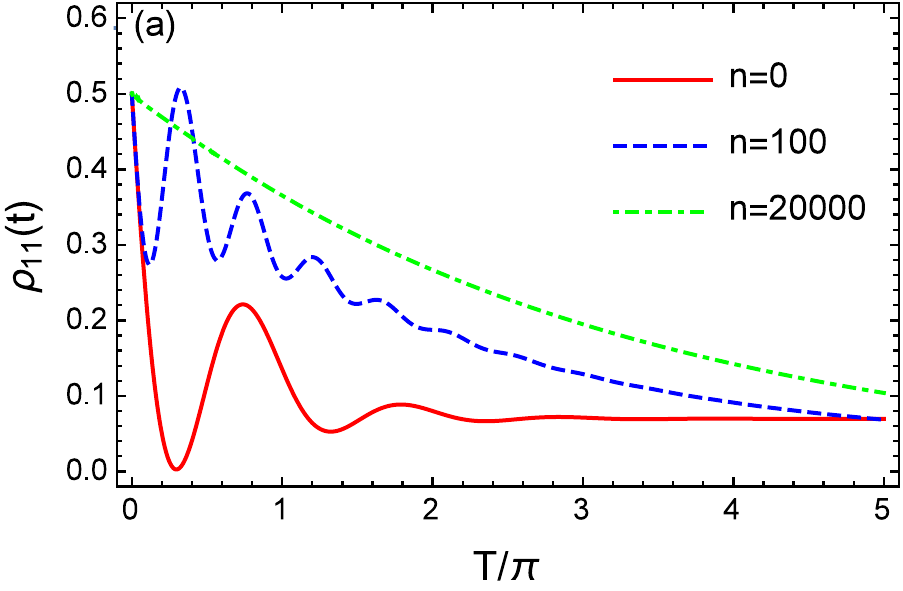}
\centering\includegraphics[width=7.5cm]{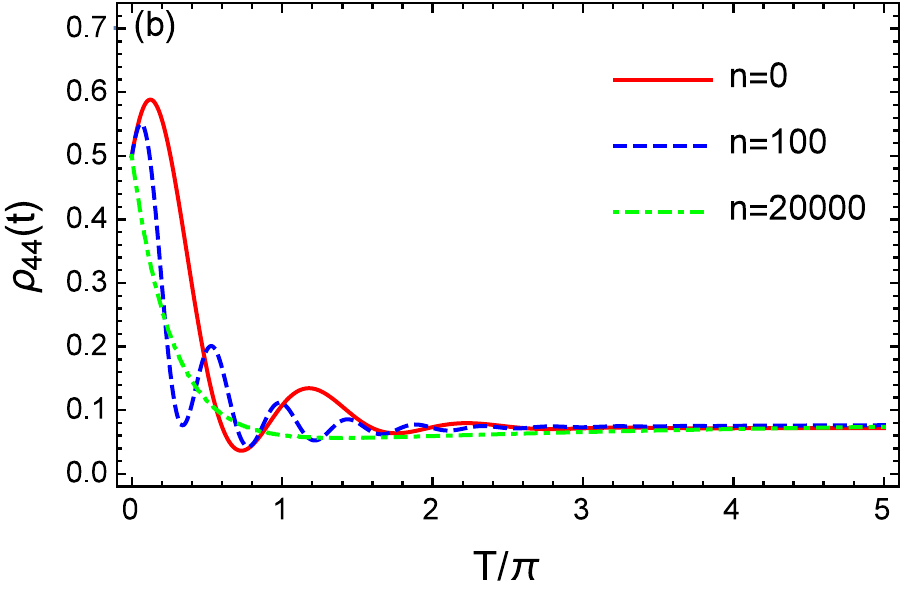}
\centering\includegraphics[width=7.5cm]{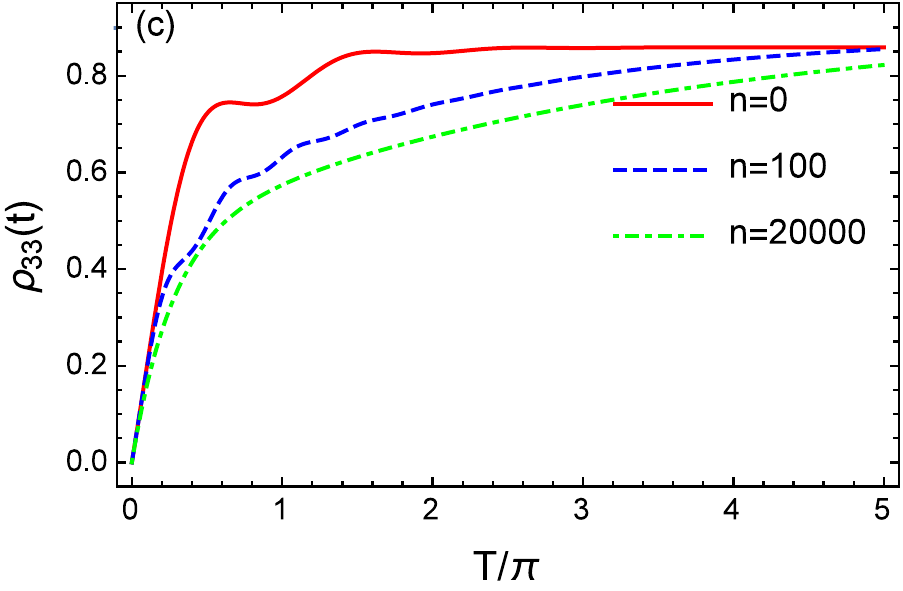}
\centering\includegraphics[width=7.5cm]{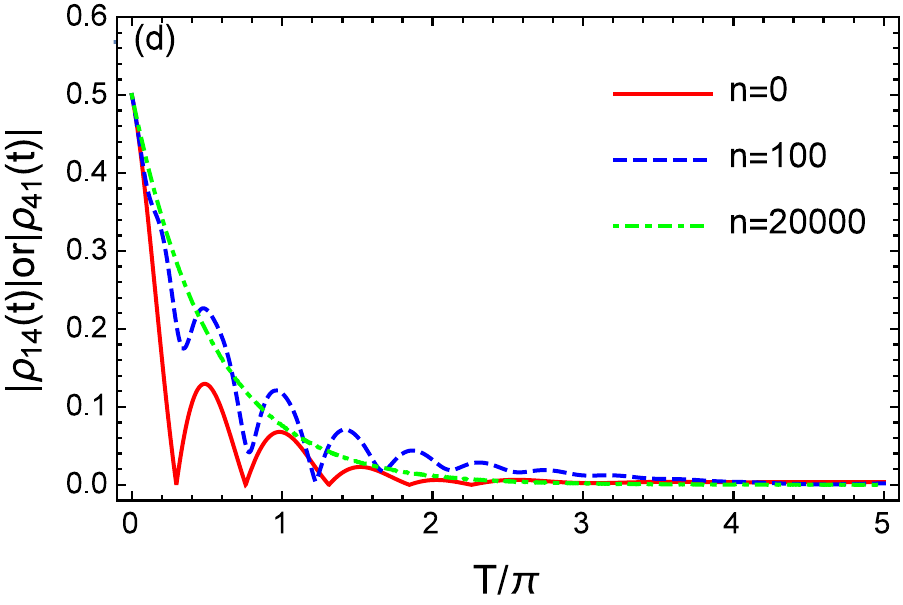}
\caption{The time evolution of diagonal and non-diagonal elements of the system density matrix
$\rho(t)$ under different photon numbers. The scaled time $T=\kappa t$,
$\gamma_{a}=0.1\kappa$, $T_{1}=0.25\pi$, $n_{th}=0.1$, and other parameters are the same as in
Fig.~2. }
\end{figure}

From Fig.~4, under considering the interaction between the environment and the system, one can
observe that the diagonal elements $\rho_{11}$ and $\rho_{44}$ evolve with time periodically, but
their amplitudes gradually decrease. On the contrary, the amplitude of diagonal element
$\rho_{33}$ increases first and then keeps a certain value. This implies that the quantum state of system
decays from the entangled state of $|\downarrow, 1\rangle$ and $|\uparrow, 0\rangle$ to the ground
state $|\downarrow, 0\rangle$ gradually, accompanied by energy decaying from the system to the
environment. Besides, the non-diagonal elements $\rho_{14}$ and $\rho_{41}$
also decays with time periodically, that is, the quantum coherence gradually disappears.
However, one can also find that through increasing the photon number in the ancillary cavity, the 
transition from the entangled state to the ground state can be effectively mitigated (Fig.~4(c)),
meanwhile, the decoherence can also be slowed down (Fig.~4(d)). Furthermore, when the number of photon
is large enough, one can see that the curves of the time evolution will not oscillate anymore, which
can be understood with the photon-dependent Rabi oscillation period $t_r$. The Rabi oscillation
period $t_r$ is inversely proportional to the Rabi oscillation frequency, i.e.,
\begin{equation}
t_r\sim \frac{1}{\widetilde{\Omega}_{n}}\propto \frac{1}{n}.
\end{equation}
Thus, the period of the oscillation will tend to zero if the number of photons is large enough, that is,
the curves will evolve with time without oscillation. From the above analysis, we can see that with
controlling the photon number in the ancillary cavity, the energy loss and the decoherence of the
system can be effectively mitigated.

Now we investigate the effects of the photon on the spin-oscillator
entanglement under the environment-induced dissipation and decoherence. Due to the 
dissipation and the decoherence, the system quantum state is a mixed
state. Based on the definition of concurrence for the mixed state \cite{wootters1998entanglement},
we numerically simulate the time evolution of the concurrence with different photon 
numbers and thermal phonon numbers, as shown in Fig.~5. From the curves, one can observe that 
the concurrence evolves with time periodically, but its amplitude
gradually decays to zero. This is because in the time evolution with dissipation and decoherence,
the system quantum state gradually decays from the entangled state of $|\downarrow, 1\rangle$
and $|\uparrow, 0\rangle$ to the ground state $|\downarrow, 0\rangle$, meanwhile the system
quantum coherence also decreases with time, which can be seen from Fig.~4. However, through
increasing the photon number in the ancillary cavity, we can increase the time and the degree of
the spin-oscillator entanglement. Besides, when the number of photon is large enough, one can 
also see that the cure of the concurrence will not oscillate anymore, which is due to the decrease of
oscillation period $t_r$. Furthermore, from Fig.~5, one can find that, through increasing the photon
number, the generated spin-oscillator entanglement is robust to the environment temperature.
\begin{figure}[h!]
\centering\includegraphics[width=7.5cm]{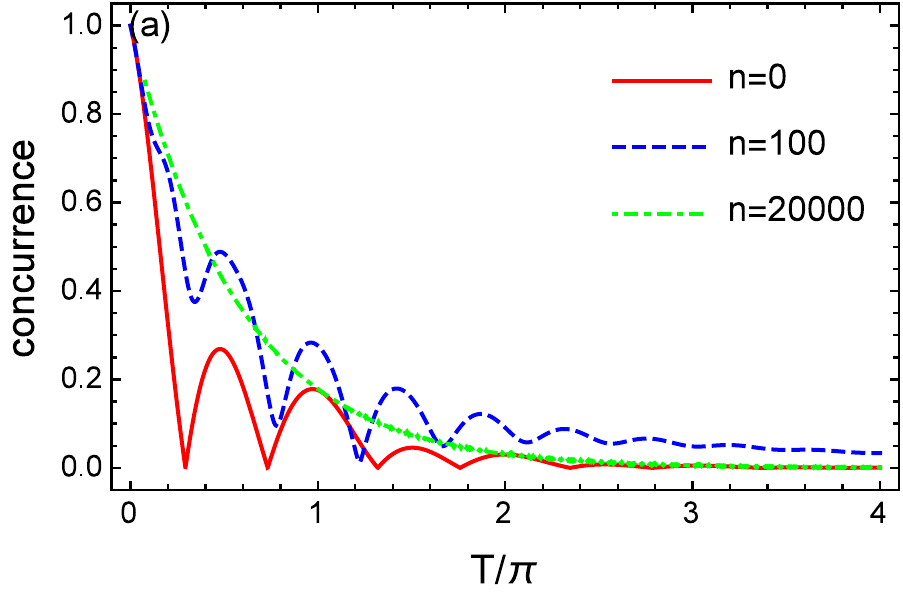}
\centering\includegraphics[width=7.5cm]{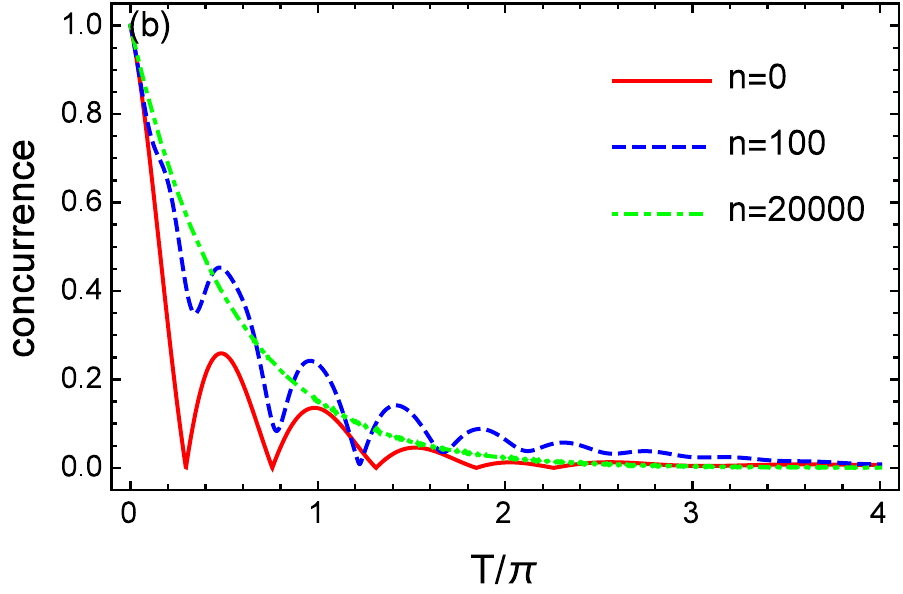}
\caption{The time evolution of the concurrence with different photon numbers and thermal phonon
numbers under considering the environment-induced decoherence and dissipation. The parameters
are: (a) $n_{th}=0$, (b) $n_{th}=0.1$, and other parameters are the same as in Fig.~4.}
\end{figure}

From the above analysis, one can find that through manipulating the number of photon in the ancillary cavity, a fully switchable
spin-oscillator entanglement is realized, and under considering the environment-induced decoherence and
dissipation, the time and the degree of entanglement can also be increased.

\section{\label{sec:level4}photon-assisted mechanical squeezing}
In this section, we investigate the situation where the detuning $\Delta$ between the spin and the
oscillator is large enough, specifically, 
$\Delta\gg\lambda_{n}\sqrt{\left\langle b^{\dagger}b\right\rangle }$. Then, an effective Hamiltonian
for the interaction Hamiltonian (i.e., Eq.~(5)) can be derived as \cite{gerry2005introductory}
\begin{equation}
H_{eff}^{\prime}=\chi \left[\left(b^{\dagger}b+1\right)|\uparrow \rangle\left\langle \uparrow \left|-b^{\dagger}b\right|\downarrow \right\rangle \langle \downarrow|\right]
\end{equation}
with $\chi=\frac{\lambda_{n}^2}{\Delta}$.
Projecting the effective Hamiltonian $H_{eff}^{\prime}$ into the spin-up subspace,  Eq.~(18) becomes
\begin{align}
H_{\uparrow}&=\left\langle \uparrow |H_{eff}^{\prime}|\uparrow \right\rangle \notag \\
&=\chi b^{\dagger}b,
\end{align}
in which the constant term has been neglected.

From Eq.~(19), one can see that, under the large detuning between the spin and the oscillator, the
oscillator can be decoupled with the spin, and is dependent on the photon of the ancillary
cavity and the environment. Considering the dissipation induced by the oscillator-bath coupling, the
quantum Langevin equation for the mechanical mode $b$ can be derived as \cite{breuer2002theory}
\begin{equation}
\dot{b}=-(\kappa+i\chi)b+\sqrt{2\kappa}b_{in},
\end{equation}
in which $b_{in}$ is the noise operator for the thermal bath, and satisfies the following nonzero
correlation functions,
\begin{align}
\left\langle b_{in}^{\dagger}(t)b_{in}(t^{\prime})\right\rangle &=2\pi n_{th} \delta (t-t^{\prime}),\\
\left\langle b_{in}(t)b_{in}^{\dagger}(t^\prime)\right\rangle &=2\pi (n_{th}+1) \delta (t-t^{\prime}).
\end{align}
Besides, the noise operator $b_{in}$ has a zero-mean value, that is, 
$\left\langle b_{in}\right\rangle=0$.

Then, Eq.~(20) can be solved analytically by the Laplace transform as follows \cite{sete2010interaction,zhang2018quantum},
\begin{equation}
b(t)=f(t)b(0)+\sqrt{2\kappa}\int_{0}^{t}f(t-t^{\prime})b_{in}(t^{\prime})dt^{\prime}
\end{equation}
with
\begin{equation}
f(t)=\textrm{exp}[-(\kappa+i\chi)t].
\end{equation}

Now we analyze the effects of the photon and the dissipation on the squeezing properties of the
oscillator. The squeezing of the oscillator can be evaluated by the variances of its quadrature
operators, $X_{+}=S(r_{n})(b^{\dagger}+b)S^{\dagger}(r_{n})$ and
$X_{-}=S(r_{n})[i(b^{\dagger}-b)]S^{\dagger}(r_{n})$, as follows \cite{sete2010interaction,zhang2018quantum},
\begin{align}
\left\langle \text{\ensuremath{\Delta X_{\pm}^{2}}}\right\rangle (t)	&=[1+2\left\langle b^{\dagger}b\right\rangle \pm(\left\langle b^{2}\right\rangle +\left\langle b^{\dagger2}\right\rangle ) \notag \\
&\mp(\left\langle b^{\dagger}\right\rangle \pm \left\langle b\right\rangle )^{2}]e^{\pm2r_{n}} \notag \\
&=[1+2n_{th}(1-e^{-2\kappa t})]e^{\pm2r_{n}},
\end{align}
in which the oscillator is considered to be initially in the vacuum state, that is, 
\begin{equation}
\left\langle b^{\dagger}(0)b(0)\right\rangle =0.
\end{equation}

From Eq.~(25), one can find that the squeezing of the oscillator can only occur in 
$\left\langle \text{\ensuremath{\Delta X_{-}^{2}}}\right\rangle (t)$, and there is a steady-state
variance, i.e.,
\begin{equation}
\left\langle \text{\ensuremath{\Delta X_{-}^{2}}}\right\rangle _{ss}=(1+2n_{th})e^{-2r_{n}},
\end{equation}
as shown in  Fig.~6(a).

In order to further analyze the role of photon number in the generation of the mechanical
squeezing, we also plot the steady-state variance 
$\left\langle \text{\ensuremath{\Delta X_{-}^{2}}}\right\rangle_{ss}$ as a function of the
photon number $n$ for different thermal phonon numbers $n_{th}$, as shown in Fig.~6(b). One
can obviously observe that the squeezing can be effectively manipulated by the photon. Specifically,
when there are no photons in the ancillary cavity, there is no squeezing in the mechanical mode.
However, through increasing the photon number, the mechanical squeezing occurs and the
squeezing degree can be optimized. For example, when the photon number $n=24000$, the
steady-state variance $\left\langle \text{\ensuremath{\Delta X_{-}^{2}}}\right\rangle_{ss}\approx 0.2$.
What's more, as long as the condition for the RWA is not broken, the mechanical mode can be
further squeezed by continuously increasing the photon number, for example, when
$n=24800$, $\left\langle \text{\ensuremath{\Delta X_{-}^{2}}}\right\rangle_{ss} \approx 0.09$.
Besides, one can also find that the generated mechanical squeezing is robust to the environment
temperature.
\begin{figure}[h!]
\centering\includegraphics[width=7.5cm]{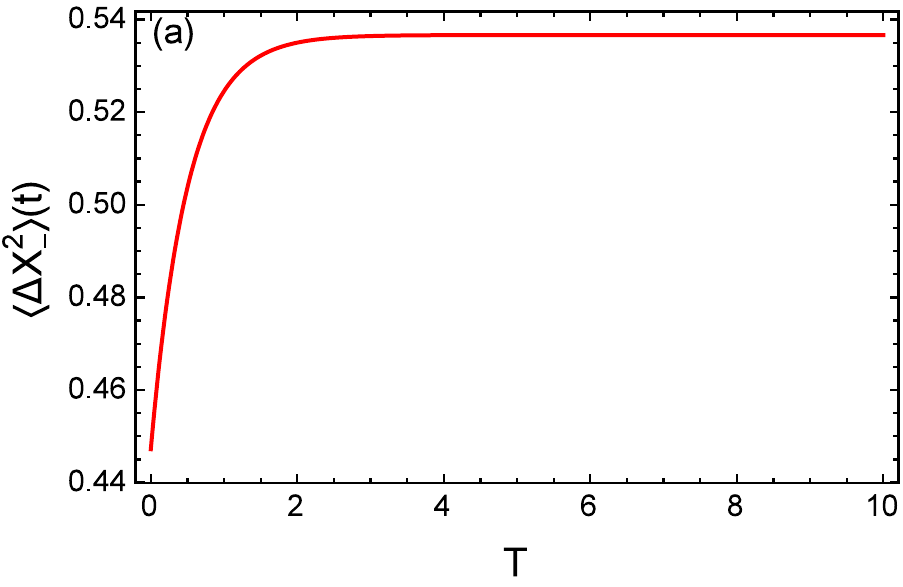}
\centering\includegraphics[width=7.5cm]{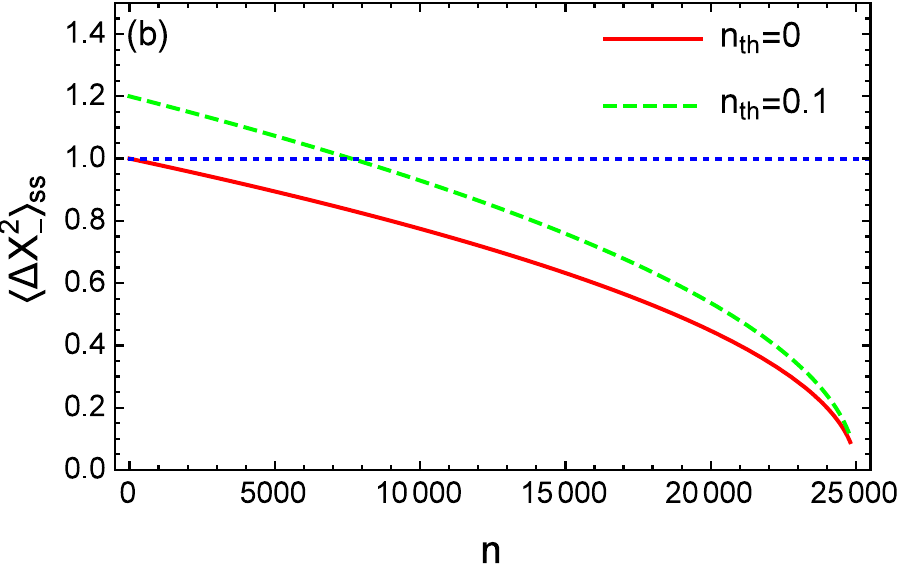}
\caption{(a) The time evolution of the time-dependent variance
$\left\langle \text{\ensuremath{\Delta X_{-}^{2}}}\right\rangle (t)$. (b) Plot of the steady-state
variance $\left\langle \text{\ensuremath{\Delta X_{-}^{2}}}\right\rangle_{ss}$ as a function of the
photon number $n$ for different thermal phonon numbers $n_{th}$, where above the dotted blue
line, there is no squeezing in the mechanical mode, and below the line, there is squeezing.
The parameters are: (a) $T=\kappa t$, $n=20000$, $n_{th}=0.1$; (b) $n_{th}=0,$ $0.1$, and other
parameters are the same as in Fig.~2.}
\end{figure}

\section{\label{sec:level5}Conclusion}
In conclusion, we have studied spin-oscillator entanglement and mechanical squeezing in the open spin-optomechanical system.
We showed that the spin-oscillator coupling and detuning can be effectively modulated by the photons of the
ancillary cavity, which leads to the generation of a fully switchable spin-oscillator entanglement and a
strong mechanical squeezing. 

Besides, we also showed that under considering the environment-induced decoherence 
and dissipation, the entanglement time and degree between spin and oscillator can be significantly improved
by increasing the number of photons, meanwhile a robust mechanical squeezing can also be generated.
This work realizes entanglement and quantum squeezing, and decoherence suppression in
the open quantum system, which has potential applications in the quantum technologies.

\section*{Acknowledgment}
This work was supported by the National Key Research and Development Program of China (Grants No.~2017YFA0304202 and
No.~2017YFA0205700), the NSFC (Grants No.~11875231 and No.~11935012), 
and the Fundamental Research Funds for the Central Universities through Grant No.~2018FZA3005.

\section*{Appendix: Solution of the Schr{\"o}dinger equation}
By substituting Eqs.~(5)-(6) into the Schr{\"o}dinger equation, one can get the following differential
equations on the probability amplitudes,
\begin{align}
\frac{d{{c}_{\uparrow ,\text{0}}}}{dt_1}&=-i\frac{{{\Omega }_{n}}}{2}{{e}^{i\Delta t_1}}{{c}_{\downarrow,1}},\\ 
\frac{d{{c}_{\downarrow ,1}}}{dt_1}&=-i\frac{{{\Omega }_{n}}}{2}{{e}^{-i\Delta t_1}}{{c}_{\uparrow,0}},
\end{align}
with $\Omega_{n}=2\lambda_{n}$.

In order to eliminate the time factors (i.e., $e^{-i\Delta t_1}$ and $e^{i\Delta t_1}$) in the above equations,
one can replace them with the following new variables,
\begin{align}
{{\tilde{c}}_{\uparrow ,0}}\left( t_{1} \right)&={{c}_{\uparrow ,0}}\left( t_{1} \right){{e}^{-i\frac{\Delta }{2}t_{1}}},\\
{{\tilde{c}}_{\downarrow ,1}}\left( t_{1} \right)&={{c}_{\downarrow ,1}}\left( t_{1} \right){{e}^{i\frac{\Delta }{2}t_{1}}}.
\end{align}
Then we have following differential equations for the new variables,
\begin{align}
\frac{d}{dt_{1}}{{\tilde{c}}_{\uparrow ,0}}&=-i\frac{\Delta }{2}{{\tilde{c}}_{\uparrow ,0}}-i\frac{{{\Omega }_{n}}}{2}{{\tilde{c}}_{\downarrow ,1}},\\
\frac{d}{dt_{1}}{{\tilde{c}}_{\downarrow ,1}}&=-i\frac{{{\Omega }_{n}}}{2}{{\tilde{c}}_{\uparrow ,0}}+i\frac{\Delta }{2}{{\tilde{c}}_{\downarrow ,1}}.
\end{align}
These equations don't have time factors, which can be solved as,
\begin{align}
{{\tilde{c}}_{\uparrow ,0}}\left( t_{1} \right)&=\cos \left( \frac{1}{2}{{{\tilde{\Omega }}}_{n}}t_{1} \right)-i\frac{\Delta }{{{{\tilde{\Omega }}}_{n}}}\sin \left( \frac{1}{2}{{{\tilde{\Omega }}}_{n}}t_{1} \right),\\
{{\tilde{c}}_{\downarrow ,1}}\left( t_{1} \right)&=-i\frac{{{\Omega }_{n}}}{{{{\tilde{\Omega }}}_{n}}}\sin \left( \frac{1}{2}{{{\tilde{\Omega }}}_{n}}t_{1} \right),
\end{align}
with $\widetilde{\Omega}_{n}=\sqrt{\Omega_{n}^{2}+\Delta^2}$. Thus, substituting the above results
into Eqs.~(30)-(31), one can get the solution of the Schr{\"o}dinger equation.
\bibliography{Manuscript}% Produces the bibliography via BibTeX.

\end{document}